\documentclass[pra,aps,twocolumn,longbibliography,preprintnumbers,superscriptaddress,floatfix]{revtex4-2}
\usepackage[dvipsnames]{xcolor}
\usepackage[left=1.2cm,right=1.2cm,top=1.2cm,bottom=1.2cm]{geometry}
\definecolor{mediumpersianblue}{rgb}{0.0, 0.4, 0.65}
\definecolor{persianred}{rgb}{0.8, 0.2, 0.2}
\definecolor{persianorange}{rgb}{0.85, 0.56, 0.35}
\definecolor{timberwolf}{rgb}{0.86, 0.84, 0.82}
\usepackage[colorlinks=true,citecolor=persianred,linkcolor=mediumpersianblue, urlcolor=mediumpersianblue]{hyperref}
\usepackage{physics}
\usepackage{graphicx}
\usepackage{amsmath}
\usepackage{amssymb}
\usepackage{bm}
\usepackage{xspace,slashed}
\usepackage{enumitem}
\usepackage{algorithm2e}
\SetKwComment{Comment}{/* }{ */}
\usepackage{multirow}
\usepackage[utf8]{inputenc}
\usepackage{bm}
\usepackage{braket}
\usepackage{array}
\usepackage{tikz}
\usepackage{fontawesome}
\usetikzlibrary{quantikz}

\usepackage{xspace,slashed}
\usepackage{array}
\usepackage{fixme}

\newcommand{\coloredcircle}[1]{%
  \tikz\draw[line width=0.25mm, black, fill=#1] (0,0) circle (0.15cm);%
}

\usepackage{listings}
\usepackage{xcolor}
\definecolor{codegreen}{rgb}{0,0.6,0}
\definecolor{codegray}{rgb}{0.5,0.5,0.5}
\definecolor{codepurple}{rgb}{0.58,0,0.82}
\definecolor{backcolour}{rgb}{0.95,0.95,0.92}
\definecolor{lightgreen}{rgb}{0.56, 0.93, 0.56}
\definecolor{lightkhaki}{rgb}{0.94, 0.9, 0.55}
\definecolor{unitednationsblue}{rgb}{0.36, 0.57, 0.9}

\lstdefinestyle{mystyle}{
    backgroundcolor=\color{backcolour},
    commentstyle=\color{persianred},
    keywordstyle=\color{persianorange},
    numberstyle=\tiny\color{codegray},
    stringstyle=\color{mediumpersianblue},
    basicstyle=\ttfamily\footnotesize,
    breakatwhitespace=false,
    breaklines=true,
    captionpos=b,
    keepspaces=true,
    numbers=left,
    numbersep=5pt,
    showspaces=false,
    showstringspaces=false,
    showtabs=false,
    tabsize=2
}

\usepackage{csquotes}
\usepackage[caption=false]{subfig}
\newcommand{\Qibo}{\texttt{Qibo}\xspace}
\newcommand{\Qiboml}{\texttt{Qiboml}\xspace}
\newcommand{\Qibolab}{\texttt{Qibolab}\xspace}
\newcommand{\Qibocal}{\texttt{Qibocal}\xspace}

\newcommand{\QiboTN}{\texttt{QiboTN}\xspace}

\newcommand{\TensorFlow}{\texttt{TensorFlow}\xspace}
\newcommand{\Keras}{\texttt{Keras}\xspace}
\newcommand{\PyTorch}{\texttt{PyTorch}\xspace}
\newcommand{\Jax}{\texttt{Jax}\xspace}
\newcommand{\PennyLane}{\texttt{PennyLane}\xspace}
\newcommand{\NumPy}{\texttt{NumPy}\xspace}

\newcommand{\eg}{\emph{e.g.}\xspace}

\fxsetup{status=draft, layout=inline, theme=color}
\definecolor{fxnote}{rgb}{1.000,0.0000,0.0000}
\usepackage{utfsym}

\begin{document}
\title{\texttt{Qiboml}: towards the orchestration of quantum-classical machine learning}

\preprint{TIF-UNIMI-2025-20}

\newcommand{\MIaff}{Dipartimento di Fisica, Universit\`a degli Studi di Milano, Milan, Italy}
\newcommand{\INFN}{INFN, Sezione di Milano, I-20133 Milan, Italy}
\newcommand{\TII}{Quantum Research Center, Technology Innovation Institute, Abu Dhabi, UAE}
\newcommand{\CERNaff}{European Organization for Nuclear Research (CERN), Geneva 1211, Switzerland}
\newcommand{\ANU}{School of Computing, The Australian National University, Canberra, ACT, Australia}
\newcommand{\Boston}{Faculty of Computing \& Data Sciences, Boston University, Boston, MA, USA}
\newcommand{\Sap}{Dipartimento di Fisica, Universit\`a la Sapienza, Rome, Italy}
\newcommand{\NTU}{Division of Physics and Applied Physics, School of Physical and Mathematical Sciences, Nanyang Technological University, Singapore}
\newcommand{\HelTeq}{QTF Centre of Excellence, Department of Physics, University of Helsinki, FI-00014 Helsinki, Finland}

\author{Matteo Robbiati}
\thanks{Equal contribution.}
\affiliation{\MIaff}
\affiliation{\CERNaff}

\author{Andrea Papaluca}
\thanks{Equal contribution.}
\affiliation{\MIaff}
\affiliation{\INFN}
\affiliation{\ANU}

\author{Andrea Pasquale}
\affiliation{\MIaff}
\affiliation{\INFN}

\author{Edoardo Pedicillo}
\affiliation{\MIaff}
\affiliation{\TII}

\author{Renato M. S. Farias}
\affiliation{\TII}

\author{Alejandro Sopena}
\affiliation{\TII}

\author{Mattia Robbiano}
\affiliation{\HelTeq}

\author{Ghaith Alramahi}
\affiliation{\TII}
\affiliation{\Boston}

\author{Simone Bordoni}
\affiliation{\TII}
\affiliation{\Sap}

\author{Alessandro Candido}
\affiliation{\TII}

\author{Niccolò Laurora}
\affiliation{\MIaff}

\author{Jogi Suda Neto}
\affiliation{\CERNaff}

\author{Yuanzheng Paul Tan}
\affiliation{\NTU}

\author{Michele Grossi}
\affiliation{\CERNaff}

\author{Stefano Carrazza}
\affiliation{\MIaff}
\affiliation{\INFN}
\affiliation{\TII}

\begin{abstract}
We present \Qiboml, an open-source software library for orchestrating
quantum and classical components in hybrid machine learning workflows. Building
on \Qibo's quantum computing capabilities and integrating with popular machine
learning frameworks such as \TensorFlow and \PyTorch, \Qiboml enables the
construction of quantum and hybrid models that can run on a broad range of
backends: (\emph{i}) multi-threaded CPUs, GPUs, and multi-GPU systems for
simulation with statevector or tensor network methods; (\emph{ii}) quantum
processing units, both on-premise and through cloud providers. In this paper, we
showcase its functionalities, including diverse simulation options, noise-aware
simulations, and real-time error mitigation and calibration.
\end{abstract}

\maketitle

\tableofcontents

\section{Introduction}
\label{sec:introduction}

Quantum machine learning (QML) investigates how quantum information processing can
be combined with learning objectives~\cite{Schuld_2014, Biamonte_2017}. A central line
of research focuses on parametrized quantum circuits (PQCs), which serve as models to
prepare quantum states. From these states, classical statistics are extracted to support
downstream tasks such as classification, regression, generative modeling, and control~\cite{Cerezo_2021}.
From a computational complexity perspective, it remains an open
problem to determine when such models can provide practical advantages over the leading
classical approaches. Nonetheless, identifying possible regimes where quantum models may be
beneficial and the architectural features that make them trainable and robust is a
necessary step for progress in both quantum computing and machine learning.

\begin{figure*}[ht]
	\center
	\includegraphics[width=0.8\textwidth]{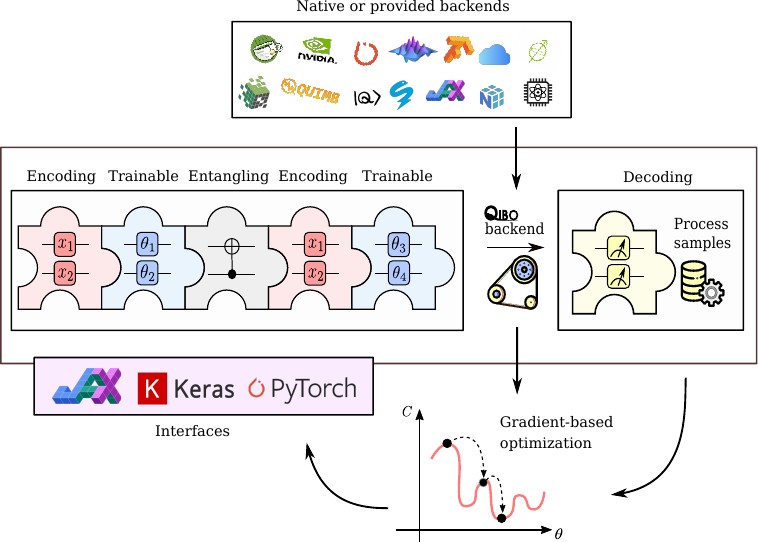}
	\caption{\label{fig:qiboml_scheme} Schematic representation of a quantum
		machine learning pipeline with \Qiboml.}
\end{figure*}

Rather than being in competition, classical and quantum routines should be viewed as
complementary. On the one hand, they can be combined to design hybrid
algorithms in which quantum circuits are embedded within broader classical workflows.
On the other hand, in the long term, quantum processors are expected to play the role of
specialized accelerators within large-scale computing infrastructures, in a way analogous
to how GPUs are employed today.
Hybrid algorithms, in which a classical optimizer updates the circuit
parameters using information from quantum measurements, are currently the standard training
approach and represent a consolidated interface between the two paradigms~\cite{Cerezo_2021}.
In many of the most promising QML applications, quantum
computers appear as subroutines within broader classical workflows, for example, in data
feature extraction or in generating samples for hybrid architectures~\cite{Bravo_Prieto_2022,Belis_2024}.
Combining classical techniques with quantum machine learning can also mitigate some main
limitations of current quantum devices, such as noise and trainability issues~\cite{McClean_2018,
crognaletti2025}.

At the same time, classical machine learning remains the reference technology for a wide
range of tasks and will continue to dominate in the foreseeable future. The relevant
question is therefore not whether quantum models replace classical ones, but how quantum
components can be integrated into established workflows in a reproducible, modular, and
tool-compatible way. Compatibility with widely used machine learning frameworks is
crucial to make QML accessible to both academic and non-academic communities.

With this motivation, we introduce \Qiboml \texttt{0.1.0}~\cite{qiboml_code}, an open-source library for quantum machine
learning developed within the \Qibo ecosystem \cite{Efthymiou2021, Carrazza_2023}. The package has two main objectives:
\textit{(i)} to provide a concise interface for building and training quantum and hybrid
models in mainstream machine learning (ML) environments, exposing quantum layers and decoders that
behave like standard \TensorFlow \cite{Abadi2016} and \PyTorch \cite{Paszke2019} modules; and \textit{(ii)} to offer
full-stack control over QML workloads, from high-level model definitions to pulse-level
scheduling on self-hosted devices, enabling end-to-end experimentation under a single
open-source stack. Figure~\ref{fig:qiboml_scheme} illustrates how these two aspects
combine in a typical \Qiboml workflow.

These goals are enabled by the design of \Qibo, which provides a unified front end for
circuit construction and a collection of interchangeable backends for simulation and
hardware execution, including differentiable simulators and laboratory control 
layers~\cite{Efthymiou2021,Efthymiou2024,Pasquale2024, qibocal_paper}.
Within this environment, \Qiboml adds the QML-specific abstractions required to (\emph{i}) encode data,
(\emph{ii}) combine trainable quantum blocks with classical layers, (\emph{iii}) decode
measurement outcomes into losses and metrics, and (\emph{iv}) integrate with automatic
differentiation and optimizers from the host ML framework. This design allows users to
prototype and train a model on a local simulator, switch to a tensor network or \emph{just-in-time} (JIT)
accelerated backend for larger scale~\cite{Efthymiou_2022}, and eventually deploy on hardware, without
modifying the high-level code.

Several existing libraries provide parts of this functionality, including \PennyLane \cite{Bergholm2022} and
\TensorFlow~\texttt{Quantum}~\cite{broughton2021}. \Qiboml differs in
scope: it is natively integrated with \Qibo's simulators and hardware middleware,
providing a reproducible workflow from model definition to pulse-level execution and
calibration, entirely within open-source components. This makes \Qiboml both a research
platform for algorithm development and an engineering tool for full-stack,
hardware-in-the-loop QML studies.

In summary, \Qiboml delivers: a consistent API for quantum layers, encoders, and
decoders that integrate with standard ML training loops; compatibility with multiple
simulation backends and automatic differentiation; and a direct bridge to laboratory
execution through \Qibo’s middleware. Together, these elements establish a practical
foundation for studying hybrid learning workflows and benchmarking quantum models
against classical baselines within the same software stack~\cite{Cerezo_2021,Efthymiou2021,Efthymiou2024,Pasquale2024,Bergholm2022,broughton2021}.

The remainder of the paper is organized as follows. Section~\ref{sec:structure} describes the
package design and its main features. Section~\ref{sec:using_qiboml} presents a series of
experiments, including a regression task, a variational quantum eigensolver (VQE)
example, and a study of noise modeling and mitigation strategies. In
Section~\ref{sec:vs_pennylane}, we benchmark \Qiboml against state-of-the-art libraries.
Finally, Section~\ref{sec:conclusion} summarizes the work and outlines future
developments.

%%%%%%%%%%%%%%%%%%%%%%%%%%%%%%%%%%%%%%%%%%%%%%%%%%%%%%%
\section{Software design}
\label{sec:structure}
%%%%%%%%%%%%%%%%%%%%%%%%%%%%%%%%%%%%%%%%%%%%%%%%%%%%%%%

As highlighted in Figure \ref{fig:qiboml_scheme}, the philosophy of \Qiboml is to
remain fully transparent to the ML interface, while leveraging the quantum
backend provided by \Qibo.
This ensures that quantum models can be trained as seamlessly as classical ones,
benefiting from optimization, differentiation, and model management tools
already available in frameworks such as \Keras~\cite{Chollet2015} and \PyTorch.

From a structural perspective, the package is organized into five main
modules: (\emph{i}) models building blocks, which include encodings, ansätze,
and decodings; (\emph{ii}) interfaces with ML frameworks, which wrap the
quantum models as native \Keras or \PyTorch objects; (\emph{iii}) computational backends
that support automatic differentiation, (\emph{iv}) custom differentiation engines for broader
compatibility with non-natively differentiable backends; and (\emph{v}) support
components, which orchestrate the interaction between quantum and classical
computations, including mitigation and calibration strategies.

\subsection{\Qiboml's model building blocks}

The \texttt{qiboml.models} module defines the core \emph{quantum layers}, which
can be stacked into a \texttt{circuit\_structure} list to form complete
quantum machine learning models.

\paragraph{Encodings.} Our collection of encoders provides data encoders
that map classical input vectors to quantum circuits. These classes inherit
from the abstract \texttt{QuantumEncoding} object and only require the definition
of the \texttt{\_\_call\_\_} method specifying how inputs are transformed into sequences of
quantum gates. A canonical example is the \texttt{PhaseEncoding}, which uploads
data into the phases of single-qubit rotations. Multiple encoders can be combined
to form hybrid or composite strategies.

\paragraph{Ansätze.} A full quantum model typically includes one or more blocks of
gates that do not depend on the input data. These blocks are called \emph{trainable layers}
and, in \Qiboml, can be implemented as custom \Qibo circuits. We offer a
set of predefined ansätze, to facilitate model construction. Among them, we
provide \emph{hardware efficient} and \emph{hamming-weight preserving} ansätze.
They represent the trainable part of a model and consist of parametrized layers
of quantum gates whose parameters are optimized during training.

\paragraph{Decoders.} The decoding layers specify how to extract
useful information from the quantum circuit once executed. All decoders inherit
from the abstract \texttt{QuantumDecoding} class. Among the available decoders,
we provide \texttt{Expectation}, which computes expectation values of observables,
and \texttt{Probabilities}, which returns the probabilities of measuring each
computational basis state. Custom decoders can be implemented by inheriting from
the abstract class and defining the \texttt{\_\_call\_\_} method.

Encoders and trainable layers can be combined to form a \texttt{circuit\_structure},
which composes the main body of a quantum model. In the forward pass, a unique quantum circuit is constructed by composing all the pieces defined in the \texttt{circuit\_structure} and executed through the decoder to obtain the final outcomes.
In practice, these blocks are combined as follows:

\vspace{0.3cm}
\begin{lstlisting}[language=Python]
from qibo import Circuit, gates
from qiboml.models.encoding import PhaseEncoding
from qiboml.models.decoding import Expectation
from qiboml.interfaces.pytorch import QuantumModel

# A trainable block of gates as a Qibo circuit
nqubits = 4
circ = Circuit(nqubits)
[circ.add(gates.RY(q, theta=0.)) for q in range(nqubits)]
[circ.add(gates.RZ(q, theta=0.)) for q in range(nqubits)]

# Instantiate two independent encoders
enc1 = PhaseEncoding(nqubits)
enc2 = PhaseEncoding(nqubits)

# Build the circuit structure
circuit_structure=[enc1, circ, enc2]
\end{lstlisting}
\vspace{0.3cm}

In the following, we describe how these components are chained into full models
and integrated with ML frameworks.

\subsection{Interfaces with machine learning frameworks}

To achieve native ML integration, \Qiboml provides a common API that is exposed
to both \PyTorch and \Keras. In both cases, the central object is the
\texttt{QuantumModel}, which inherits from the corresponding framework base class
(\texttt{torch.nn.Module} or \texttt{tf.keras.Model}). This ensures that quantum
models can be seamlessly inserted into classical ML pipelines and trained using
the same tools as conventional layers.

The philosophy of the interface is that a quantum model is built from a
\texttt{circuit\_structure} (encoders and trainable layers) and a decoder,
exactly as described in the previous subsection. Once defined, the
\texttt{QuantumModel} transparently exposes all parameters as trainable objects,
and integrates with optimizers, losses, and training routines without requiring
additional wrappers.

A key design choice is that the API supports not only native automatic
differentiation from \PyTorch or \TensorFlow (see Section~\ref{sec:backends}), but also custom differentiation
engines (Section~\ref{sec:custom_diff}). In this way, models can be trained
on both simulation backends and real quantum hardware, with gradients obtained
either by standard automatic differentiation, \emph{e.g.} with our \Jax-based differentiation engine (defined in Sec. \ref{sec:structure_diff} below), or the
parameter-shift rule~\cite{Schuld2019}. The interface layer takes care of injecting the chosen
gradients into the computational graph of the ML framework, so that training
remains completely transparent to the user.

In practice, the same notation and workflow apply independently of the chosen
framework, as illustrated by the following \PyTorch example:

\vspace{0.3cm}
\begin{lstlisting}[language=Python]
from qibo.hamiltonians import TFIM
from qiboml.models.ansatze import HardwareEfficient
from qiboml.models.decoding import Expectation
import qiboml.interfaces.pytorch as pt
import torch

nqubits = 2
# Define circuit and Hamiltonian as Qibo objects
circuit = HardwareEfficient(nqubits)
ham = TFIM(nqubits, h=0.5)

decoding = Expectation(
  nqubits,
  backend="any qibo backend"
)
# Instantiate the quantum model
pt_model = pt.QuantumModel(
  circuit_structure=circuit,
  decoding=decoding
)

# What follows is standard PyTorch training
optimizer = torch.optim.Adam(pt_model.parameters(), lr=0.05)

for iteration in range(100):
    optimizer.zero_grad()
    cost = pt_model()
    cost.backward()
    optimizer.step()
\end{lstlisting}
\vspace{0.3cm}

\begin{figure}[ht]
	\center
	\includegraphics[width=\columnwidth]{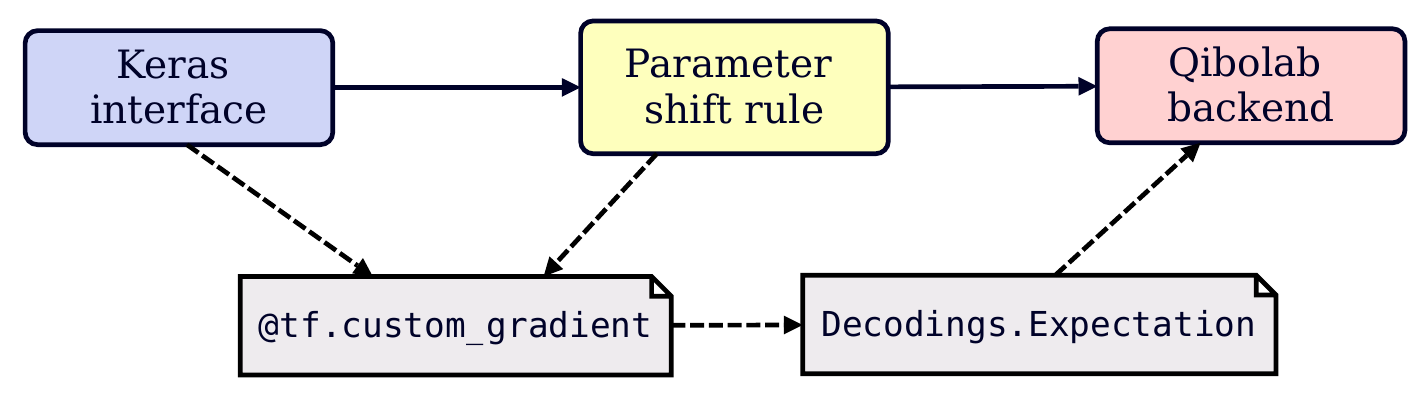}
	\caption{\label{fig:diff_example}
		Custom differentiation example using \texttt{Keras} as interface, the
		parameter-shift rule~\cite{Schuld2019} to calculate gradients, and the
		\Qibolab hardware backend to execute circuits.
	}
\end{figure}

As shown in Figure~\ref{fig:diff_example}, quantum layers appear as regular
computational, differentiable nodes in the ML framework’s graph.

\subsection{Automatically differentiable backends}\label{sec:backends}

\Qiboml is designed to be backend-agnostic, allowing users to run their models on different ML frameworks as well as on hardware platforms.
The framework abstracts away the details of the underlying computation backend and differentiation strategy,
enabling switching between different simulators and quantum devices with minimal code changes.
The main objective of the backend, in simulation, is to provide automatic differentiation capabilities,
and fasten the computation offering hardware acceleration, like GPUs and TPUs, when available.

Currently, \Qiboml supports all backends compatible with \Qibo and, in particular, relies on the \Qibolab backend for hardware execution.
\Qiboml also acts as a backend provider for \Qibo. In fact, it implements three differentiable backends,
which seamlessly integrate with \Qibo, even outside of the QML context. In particular:

\paragraph{\PyTorch.} An open-source framework developed by Meta AI, widely adopted in 
research thanks to its dynamic computation graph, ease of debugging, and strong community ecosystem.
Supporting \PyTorch ensures that \Qiboml can seamlessly integrate into the 
workflows of researchers and practitioners who already rely on it as the de facto standard in modern machine learning.

\paragraph{\TensorFlow.} An open-source framework developed by Google Brain, 
initially based on static computation graphs but later enriched with eager execution.
Its focus on scalability and production deployment makes it the framework of 
choice in many industrial applications, so supporting \TensorFlow allows \Qiboml 
to bridge research prototypes with production-ready ML pipelines and distributed large-scale training.

\paragraph{\Jax.} A high-performance numerical computing library developed by 
Google Research, with a \NumPy-like API, automatic differentiation, and JIT compilation.
\Jax is increasingly popular in scientific computing and ML research for its 
composable transformations (\texttt{grad}, \texttt{jit}, \texttt{vmap}) and 
efficient execution on GPUs/TPUs. By supporting \Jax, \Qiboml embraces a 
growing community of researchers who favor functional programming paradigms and high-performance simulations.

\subsection{Custom differentiation engines}\label{sec:custom_diff}

Since not all simulation or hardware backends natively support gradient computation, \Qiboml implements additional differentiation engines. 
These engines provide the means to calculate the Jacobian of a quantum circuit  \emph{w.r.t.} the phases of its parametrized gates and chain them with the Jacobian \emph{w.r.t.} model's parameters provided by the ML interface for seamless integration.

Among them, two \Jax-based engines enable automatic differentiation for non-natively differentiable simulation backends, but are usable in exact simulation only: 
one performing standard statevector simulation through \Jax primitives, while the other relies on \texttt{Quimb}~\cite{quimb} to execute circuits as tensor networks and, thus, allowing for the training of very large systems. 
Whereas, in the presence of sampling and, therefore, shot noise, other numerical techniques are needed. For instance, the \emph{adjoint differentiation}~\citep{jones2020} method is available in simulation and more broadly compatible to any \Qibo-like backend.
Similarly, an explicit parameter-shift rule (PSR)~\cite{Schuld2019} implementation, which is hardware-compatible, allows for gradient computation on real devices.
Finally, custom gradient strategies can also be defined by inheriting from the abstract \texttt{Differentiation} class and overloading the \texttt{evaluate} method.

\subsection{Support components}
\label{sec:structure_diff}

\Qiboml provides mechanisms for orchestrating the
quantum-classical interface, focusing on differentiation and, more generally,
gradient-aware execution of quantum models. In particular, we provide
\textit{(i)} a \texttt{CircuitTracer}, which tracks how the model's parameters
are combined to obtain the rotation angles of the gates in the circuit, thus
connecting quantum and classical parameters; and \textit{(ii)} support objects
for real-time error mitigation, real-time devices calibration, and circuit orchestration.

\paragraph{CircuitTracer.} A central component in this architecture is the
\texttt{CircuitTracer}, which provides fine-grained tracking of the high-level structure of the model.
In detail, this object traces all the operations applied to the model's parameters and the inputs to construct the complete quantum circuit during the forward pass, providing access to the Jacobian of the quantum circuit angles with respect to both the model's parameters and inputs. These Jacobians can then be chained with those provided by the custom differentiation engines (\ref{sec:structure_diff}) to obtain the complete gradient of the loss. Therefore, the \texttt{CircuitTracer} is responsible for
maintaining the classical and quantum parameters synchronized, ensuring that
derivatives are consistently injected into the ML framework.
This design makes it possible to evaluate gradients with respect to
all model parameters simultaneously, and to switch between differentiation
engines without modifying the high-level model definition.

\paragraph{Error mitigation and orchestration.} Although the current release focuses
mainly on differentiation, the architecture is designed to accommodate support
objects such as real-time error mitigators. These will coordinate the execution
of circuits and apply error-suppression techniques, building on the broader
\Qibo ecosystem. In this way, \Qiboml paves the way for reliable quantum
training workflows on noisy self-hosted hardware, without requiring any changes
to the ML interface.

\paragraph{Device calibration and pulse-level control.} Finally, \Qiboml leverages
\Qibo's middleware layer, \Qibolab, to provide direct access to self-hosted
quantum devices. This includes pulse-level control, real-time calibration, and
integration with laboratory equipment. By interfacing directly with hardware,
\Qiboml enables end-to-end QML experiments, from model definition to execution and
data acquisition, all within a single open-source stack. \\

In the following, we showcase some of these features, focusing in particular on
noise modeling and mitigation strategies. With some practical examples, we show
how to orchestrate features and interfaces inherited from \Qibo (for the quantum
computing utilities) and the host ML framework (for the classical machine learning
utilities).

\subsubsection{Quantum machine learning in a noisy setup}
\label{sec:with_noise}

\Qibo allows for the effortless construction of a noise model and \Qiboml
provides the means to easily plug it into a quantum machine learning pipeline.
Any \texttt{Decoder} accepts as an argument a \texttt{NoiseModel}, which is
then applied to the provided \texttt{circuit\_structure}.
By default, exact density matrix simulation is triggered in the presence of noise.
The following code snippet shows how to define a simple local Pauli noise model
and use it in a \Qiboml's quantum model.

\vspace{0.3cm}
\begin{lstlisting}[language=Python]
from qibo.noise import NoiseModel, PauliError
from qiboml.models.decoding import Expectation

# Building the noise model
noise_model = NoiseModel()
noise_model.add(
      PauliError(
          [
              ("X", 0.01),
              ("Y", 0.01),
              ("Z", 0.01),
          ]
      ),
      qubits=0,
  )

# Informing the decoder
# we want noisy simulation
dec = Expectation(
    nqubits=1,
    density_matrix=True,
    nshots=1024,
    noise_model=noise_model,
)
\end{lstlisting}
\vspace{0.3cm}

When \Qibolab is used to run on real devices, the noise will be the natural
noise from the quantum hardware, and the \texttt{noise\_model} argument is
not required, nor advised.

\subsubsection{Real-time error mitigation}
\label{sec:rtqem}

Several strategies exist for addressing noisy quantum devices. They can mainly be
divided into two groups: \emph{quantum error correction} (QEC) and
\emph{quantum error mitigation} (QEM).
While the first approach aims to correct the quantum computer output completely
removing the errors~\cite{Roffe2019}, the second approach typically consists in
performing post-processing routines which extract mitigated values leveraging the
knowledge we have about the existing noise~\cite{Cai2023}.
While QEC is surely the best solution, a reliable and scalable quantum computer
is required, and it is usually considered the golden standard for fault-tolerant
devices, but it is not easily achievable in the short term.
QEM, on the other hand, only allows for a rough estimation of the noiseless
values, but is already applicable today to near-term devices.

\begin{figure*}[!t]
	\center
	\includegraphics[width=0.77\textwidth]{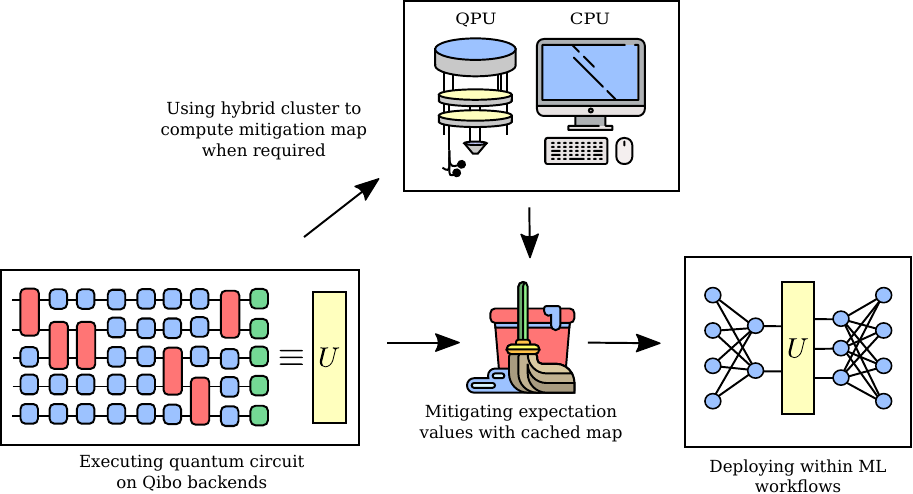}
	\caption{\label{fig:rtqem_scheme} Schematic representation of a real-time
		quantum error mitigation procedure. The mitigation map is periodically updated
		during the training, and it is used to mitigate the expectation values
		calculated for predictions and gradients. Those values are then utilized
		within the hybrid machine learning procedure.}
\end{figure*}

Among QEM techniques, in \Qiboml we focus on \textit{data-driven} methodologies.
In a nutshell, these methods consist of collecting data from noisy devices and
combining them with exact-simulated counterparts to model the noise through
classical data regressions and build a heuristic noise-inversion procedure.
A remarkable example of data-driven error mitigation is the
\emph{Clifford data regression} (CDR)~\cite{Czarnik2021}, where a family of
circuits of the same structure as the target is sampled, and, in each of these, we
replace some non-Clifford gates with Clifford ones.
This, thanks to the well-stablished stabilizer state formalism, facilitates the task
of classically simulating expectation values~\cite{Aaronson2004}.
Once the noisy data and the exact data are collected, one can fit the points and
construct a mitigation map, which is then useful to mitigate expectation values estimated using the original circuit.

In QML, it is particularly important to be able to execute circuits with the
greatest confidence possible, as they are involved in both the calculation of
the predictions of the model and in the computation of the gradients through
quantum-compatible methods, such as the parameter-shift rules.
For this reason, \Qiboml provides a real-time quantum error mitigation procedure~\cite{Robbiati2023} that takes care of gradually checking and updating a map of
the noise to be used for mitigating expectation values calculated for predictions
and gradients.
In practice, when defining a decoder, a \texttt{mitigation\_config} can be
passed as an additional argument and customized using any data-driven QEM
method implemented in \Qibo. This configuration creates an instance of a
\texttt{Mitigator} object, which is then used internally to (\emph{i}) construct
the mitigation map, (\emph{ii}) update it when it is needed, and (\emph{iii})
apply it to each expectation value calculated during the training.

As an example, we show in the following code snippet how the QEM is applied, while
an illustration of the procedure is presented in Figure \ref{fig:rtqem_scheme}.

\vspace{0.3cm}
\begin{lstlisting}[language=Python]
from qiboml.models.decoding import Expectation

# Defining the noise model as before
# Setting the QEM configuration
mitigation_config = {
	"min_iterations": 100,
	"threshold": 0.01,
	"method": "CDR",
	"method_kwargs": {"n_training_samples": 50, "nshots": 10000},
}

# Informing the decoder about the noise model
# and real-time error mitigation
dec = Expectation(
    nqubits=1,
    density_matrix=True,
    nshots=1024,
    noise_model=noise_model,
    mitigation_config=mitigation_config,
)
\end{lstlisting}
\vspace{0.3cm}

As shown, the real-time mitigation procedure can be customized through a dictionary
of parameters:
\begin{itemize}
  \item \texttt{min\_iterations} is the minimum number of expectation values,
  or decoding calls, to be computed before the cached mitigation map is \emph{checked}.
  Once checked, the map is updated only if it is considered unreliable, namely, the
  distance between a reference value and the mitigated one is above a certain threshold;
  \item \texttt{threshold} is the threshold used to decide whether the cached
  mitigation map is still reliable or needs to be updated;
  \item \texttt{method} is the error mitigation method to be used, which must be
  one of the data-driven methods implemented in \Qibo;
  \item \texttt{method\_kwargs} are additional arguments passed to the chosen
  QEM method. We suggest that the reader refer to the \Qibo documentation for
  more details on the available methods and their arguments~\cite{qibo_error_mitigation_docs}.
\end{itemize}

Through this simple interface, one can benefit from error mitigation within
the variational procedure, introducing a computational overhead regulated by
how many times the mitigation map is recomputed using the chosen technique.
The configuration chosen in this example only requires $50$ additional circuits
to be executed, while training a small model with \eg $p=20$ parameters would
require computing $40$ circuits only to estimate the gradient of the cost function once.
In practice, the overhead introduced by the real-time mitigation procedure is minimal.
Furthermore, it has been shown that even considering an evolving-noise scenario,
the number of times recomputing the mitigation map is needed can still be kept under control~\cite{Robbiati2023}.

\subsubsection{Calibration-aware training}
\label{sec:calibration}

The main challenges during the training of a quantum machine learning model
originate from the quantum hardware itself. At any timescale, the system is
subject to noise, as discussed in the previous section. Over longer time
scales, additional issues arise due to drifts in calibration parameters such as
qubit frequency and coherence times. These drifts can impede training
convergence and degrade model quality.

To address this, it is essential to continuously monitor the status of the
quantum hardware throughout training, especially in runs lasting several
hours, and recalibrate the system when necessary.

In \Qiboml, this functionality is provided by the \texttt{Calibrator}, a class
that allows users to define which \Qibocal~\cite{qibocal_paper} protocols to run and their execution
parameters. For example, one might estimate readout and qubit gate fidelities
using single-shot classification and randomized benchmarking, respectively. The
resulting data can be used either as early stopping conditions or to trigger
recalibration experiments aimed at correcting drifts. The \texttt{Calibrator}
is invoked during the execution of the \texttt{Expectation} decoder, where
users can also specify how frequently these protocols should be executed.

%%%%%%%%%%%%%%%%%%%%%%%%%%%%%%%%%%%%%%%%%%%%%%%%%%%%%%%
\section{\Qiboml in action}
\label{sec:using_qiboml}
%%%%%%%%%%%%%%%%%%%%%%%%%%%%%%%%%%%%%%%%%%%%%%%%%%%%%%%

In this section, we discuss a series of experiments to showcase \Qiboml's functionalities.
The selected targets are two: \textit{(i)} a simple regression task, to illustrate
the different training setups available; and \textit{(ii)} a variational quantum
eigensolver (VQE) example, to show how to integrate quantum models into more
complex classical workflows.

For both examples, we compare four different training setups:
\begin{itemize}
  \item[\coloredcircle{unitednationsblue}] \textbf{Noiseless and exact simulation}: the model is
  trained using \PyTorch's automatic differentiation, and the quantum
  circuit is simulated exactly, up to machine precision, without noise and with access to the full
  statevector.

  \item[\coloredcircle{lightgreen}] \textbf{Noiseless simulation with shots}: the model
  is trained without circuit noise and using a finite number of shots to estimate
  expectation values. The gradients are computed using the parameter-shift
  rule \cite{Schuld2019}.

  \item[\coloredcircle{persianred}] \textbf{Noisy simulation with shots}: the model is
  trained using a depolarising noise model and a finite number of shots to
  estimate expectation values.

  \item[\coloredcircle{lightkhaki}] \textbf{Noisy simulation with shots and real-time
  mitigation}: the model is trained using a simple noise model and a finite
  number of shots to estimate expectation values. A real-time error
  mitigation strategy is applied during the training to improve the quality
  of the results.
\end{itemize}

The idea is to show how \Qiboml can be used to easily switch between
different training setups, and can be used to orchestrate algorithmic
and hardware-oriented strategies to improve the quality of the results, not
to reach a state-of-the-art performance on the selected tasks.
%%%%%%%%%%%%%%%%%%%%%%%%%%%%%%%%%%%%%%%%%%%%%%%%%%%%%%%
\subsection{Showcasing \Qiboml's training setups}
\label{sec:benchmarks}
%%%%%%%%%%%%%%%%%%%%%%%%%%%%%%%%%%%%%%%%%%%%%%%%%%%%%%%

\begin{figure}[!ht]
	\center
	\includegraphics[width=\columnwidth]{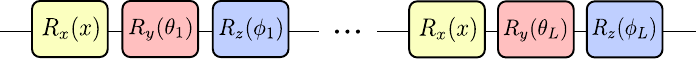}%
	\caption{\label{fig:1q_reuploading} Parametric circuit composed of $L$ layers
		of rotations. $R_x$ gates are used to encode the data (\Qiboml's encoders),
		while $R_y$ and $R_z$ gates are used as trainable gates.}
\end{figure}

As a simple one-dimensional regression example, we aim at approximating the following function:
\begin{equation}
	\label{eq:target}
	f(x) = \sin^{2}(x) - 0.3 \, \cos(x)\;.
\end{equation}
In particular, we implement the data reuploading \cite{PerezSalinas2020} model
shown in Figure \ref{fig:1q_reuploading} for a single qubit, leveraging our \texttt{QuantumModel}'s
modular structure.

\begin{figure}[!ht]
	\center
	\includegraphics[width=0.45\textwidth]{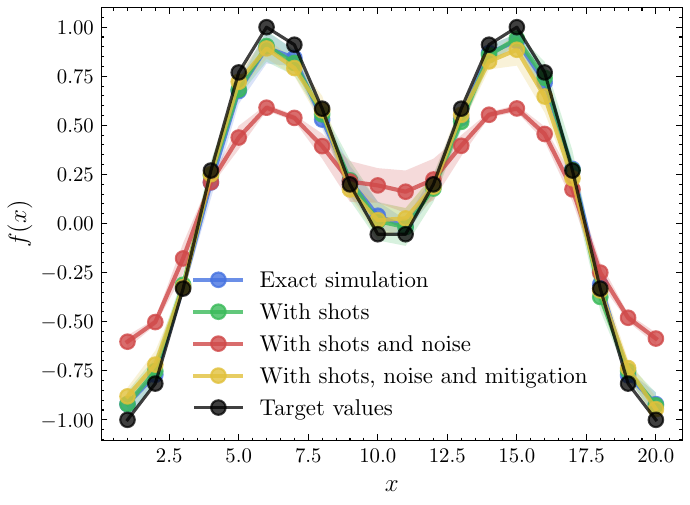}%
	\caption{\label{fig:simulations}
		Four trainings are performed with the same initial configuration shown in
		Table~\ref{tab:init_config_train}, each following a different strategy:
		noiseless and exact simulation (green), noiseless with shot-noisy simulation
		(blue), noisy with shot-noisy simulation (red), and noisy, shot-noisy with
		real-time mitigation (yellow). The approximations are compared with the
		target theoretical function introduced in Eq.~\ref{eq:target}. Solid curves and
		uncertainty intervals are obtained from the median and median absolute
		deviation of twenty repetitions, each starting from a different random seed.
	}
\end{figure}

In Figure \ref{fig:simulations}, we show four different results obtained using
classical simulations adopting the \PyTorch interface. The four trainings
correspond to the four configurations described at the beginning of this section.

In particular, infinite-shots and exact simulation is represented by the blue curve,
finite-shots and exact simulation by the green curve, finite-shots and noisy
simulation by the red curve, and finite-shots, noisy simulation with real-time
error mitigation by the yellow curve.

Noise is implemented following the procedure described in Sec. \ref{sec:with_noise}.
In this context, after each gate we apply a depolarising noise channel with depolarising parameter $10^{-2}$.

Real-time error mitigation is implemented following the procedure described in
Section \ref{sec:rtqem}, where we apply the Clifford data regression (CDR)
method \cite{Czarnik2021} to mitigate the expectation values calculated during
training (prediction and gradients, when computed through hardware-compatible
differentiation rules).
The error mitigation configuration is set as follows:

\vspace{0.3cm}
\begin{lstlisting}[language=Python]
mitigation_config = {
	"min_iterations": 5000,
	"threshold": 0.1,
	"method": "CDR",
	"method_kwargs": {
    "n_training_samples": 100, 
    "nshots": 5000
  },
}
\end{lstlisting}
\vspace{0.3cm}

Some of the training hyperparameters are shared by all the simulations presented,
and are summarized in Table~\ref{tab:init_config_train}.

\begin{table}[!ht]
	\begin{tabular}{cccc}
		\hline \hline
		Epochs & Runs & Optimizer               & Local Pauli Error prob. \\
		\hline
		$50$   & $10$ & $\text{Adam}(\eta=0.2)$ & $0.01$                  \\
		\hline \hline
	\end{tabular}
	\caption{\label{tab:init_config_train}
		The initial configuration is shared by all the presented simulations.
		In particular, we show the number of epochs of the training, the number of
		trainings per configuration (statistics used to compute the training error),
		the chosen optimizer, and, in the last column, a parameter representing the
		probability of applying $X, Y$ and $Z$ gates in case a local Pauli noise
		channel is requested.
	}
\end{table}

The yellow curve in Figure~\ref{fig:simulations} shows that real-time error mitigation
allows to recover a good approximation of the
target function, even in the presence of noise and shot noise.

%%%%%%%%%%%%%%%%%%%%%%%%%%%%%%%%%%%%%%%%%%%%%%%%%%%%%%%
\subsection{A multi-qubit example}
\label{sec:multiq_examples}
%%%%%%%%%%%%%%%%%%%%%%%%%%%%%%%%%%%%%%%%%%%%%%%%%%%%%%%

To give an example of a multi-qubit algorithm, we present here a series of trainings of VQEs.
VQEs are variational algorithms introduced to approximate the ground state of the
target Hamiltonians~\cite{Peruzzo_2014}.
In a nutshell, they consist of iteratively updating the parameters of a
parametric quantum circuit $U(\bm{\theta})$ to minimize the expectation value of
a target Hamiltonian $H_0$ over the state prepared by $U(\bm{\theta})$.

To the pedagogical purpose of this work, we tackle here a simple problem,
consisting in approximating the energy of an $n$-qubit non-interacting Pauli-$Z$
Hamiltonian $H_{0} = - \sum_{k=1}^{n} Z_{k}$, where we set $n=3$. Later in the manuscript,
we will perform a series of performance benchmarks considering larger systems.

\begin{figure}[ht]
	\center
	\includegraphics[width=0.48\textwidth]{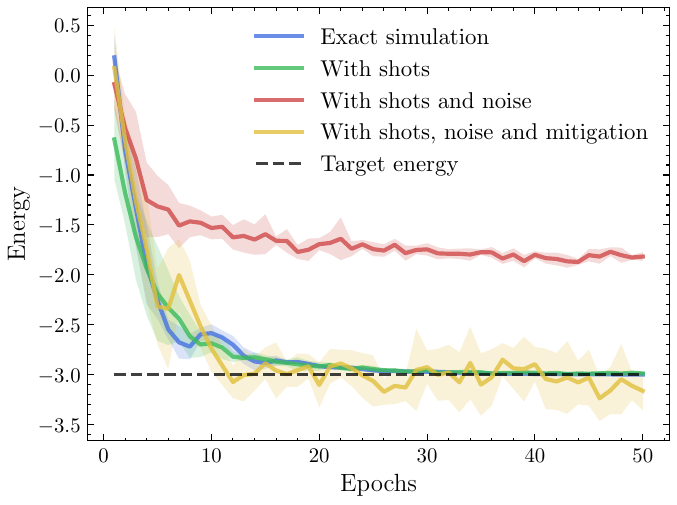}
	\caption{\label{fig:vqes} Four trainings are performed with the same initial configuration shown in
	Table~\ref{tab:init_config_train}, each following a different strategy:
	noiseless and exact simulation (green), noiseless with shot-noisy simulation
	(blue), noisy with shot-noisy simulation (red), and noisy, shot-noisy with
	real-time mitigation (yellow). The approximations are compared with the
	target ground state energy (black line). Solid lines and uncertainty
	intervals are obtained from the median and median absolute deviation of
	twenty repetitions, each starting from a different random seed.
	}
\end{figure}

We consider the same training setups introduced at the beginning of this section
and described in the one-dimensional regression example. In this case, the same
local Pauli noise channel is applied, but we set $q=0.008$. The real-time error
mitigation configuration is shown in the following code snippet.

\vspace{0.3cm}
\begin{lstlisting}[language=Python]
mitigation_config = {
	"threshold": 0.2,
	"min_iterations": 500,
	"method": "CDR",
	"method_kwargs": {
    "n_training_samples": 100, 
    "nshots": max(nshots, 10000)
  },
}
\end{lstlisting}
\vspace{0.3cm}

The four trainings are performed using the same initial configuration,
summarized in Table~\ref{tab:init_config_vqe}.

\begin{table}[!ht]
	\begin{tabular}{ccccc}
		\hline \hline
		Epochs & Runs & Optimizer               & Local Pauli Error prob. & Qubits \\
		\hline
		$50$   & $5$  & $\text{Adam}(\eta=0.1)$ & $0.01$                  & $5$    \\
		\hline \hline
	\end{tabular}
	\caption{Initial configuration shared by all the presented simulations.
		In particular, we show the number of training epochs, the number of times each
configuration is trained (used to compute training error bars), the optimizer,
the probability of the local Pauli noise channel (in case noise is present)
and the number of qubits considered.}
	\label{tab:init_config_vqe}
\end{table}

Also in this case, we see how the training procedure can benefit from real-time error mitigation.

\subsection{Training on real hardware}

Moving from simulation to execution on real quantum hardware in \Qiboml is quite straightforward and do not
involve any significant change to the code structure. Broadly speaking, a simple reset of the backend to the approriate
\Qibolab backend is enough, together with the definition of the desired transpilation pipeline.

\begin{lstlisting}[language=Python]
  from qibo import set_backend
  from qibo.transpiler import NativeGates, Passes, Unroller
  from qibo.gates import RZ, Z, CNOT, GPI2
  
  # Setting the qibolab backend
  set_backend(
    "qibolab", 
    platform="my_local_quantum_chip"
  )

  # Defining the transpilation suitable for
  # your chip: mostly the supported connectivity
  # and the gates that are natively supported
  connectivity = [
    ("0", "1"), 
    ("0", "2"), 
    ("0", "3"), 
    ("0", "4")
  ]
  native_gates = NativeGates.from_list([
    RZ, Z, CNOT, GPI2
  ])
  transpiler = Passes(
      connectivity=connectivity, 
      passes=[Unroller(native_gates)]
  )
  # Defining the qubits you want to execute on
  wire_names = ["0", "2", "3"]
  # Attaching everything to the decoder
  decoding = Expectation(
        nqubits=nqubits,
        nshots=nshots,
        transpiler=transpiler,
        wire_names=wire_names,
  )      
\end{lstlisting}

\begin{figure}
  \includegraphics[width=\linewidth]{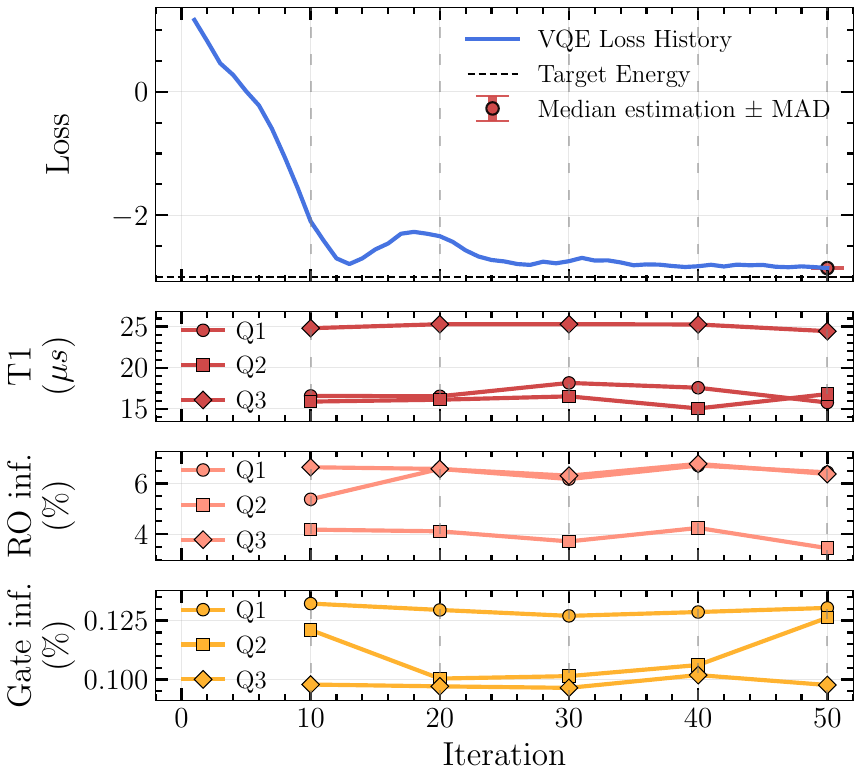}
  \caption{Ground state energy approximation training a three-qubit VQE on 
  a superconducting chip hosted in the Quantum Research Center of the 
  Technnology Innovation Institute in Abu Dhabi. A single training is performed 
  for $50$ epochs using 
  \Qiboml's \PyTorch interface. The final 
  estimation and its uncertainty are obtained as the median and median absolute deviation
  of twenty predictions computed through one thousand shots. The result is compared 
  with the exact ground state energy
  (black line). The inset plots show the coherence time $T_1$ (red), the readouts 
  and single-qubit gate infidelities (orange and yellow, respectively) tracked 
  per qubit
  during the training through a \texttt{Calibrator} object.}   
  \label{fig:vqe_tii}
\end{figure}

This simple redefinition of the decoder allows for easily testing out the previously 
introduced VQE example on a real superconducting quantum chip. As discussed in 
Section~\ref{sec:calibration}, a \texttt{Calibrator} object can be used to monitor 
the hardware status during the training. In this case, we record the coherence time $T_1$,
the readout and single-qubit gate infidelities of the three qubits used for the
training. The results of a single training are shown in Figure~\ref{fig:vqe_tii}.

% By properly defining a \texttt{Calibrator} object, the decoder 
  % was able to monitor the fidelity of single-qubit gates and readouts, as well 
  % as the coherence time $T_1$, of the three qubits throughout the training process.
  % The recorded values are shown in the inset plots and labeled accordingly.

\subsection{\label{sec:hybrid} A hybrid quantum-classical example}

\par A promising near-term avenue for quantum machine learning is the design of {\em hybrid}
algorithms in which classical and quantum components are composed into a single differentiable
pipeline~\cite{lie_eqgnn, alessandro_transformer, masha}.

Because present-day quantum processors are depth limited and noise prone, an efficient
use of the hypothesis space is paramount: one would like to leverage domain-specific inductive
biases to reduce the number of free parameters, the sample complexity, and
potentially the generalization error.
Historically, the most successful classical architectures achieve precisely this:
convolutional neural networks~\cite{deep_learning} encode approximate translational
equivariance for images, and graph neural networks (GNNs)~\cite{gnn0, gnn1, gnn2}
encode permutation invariance for molecular graphs, therefore having great interest
to the pharmaceutical industry. Some theoretical guarantees, for example, explain
how symmetry-preserving networks have better generalization bounds by learning on
a reduced subspace of orbit representatives~\cite{elesedy2022group}. The idea of
leveraging symmetries in classical neural architectures is well studied in a
broader field known as \emph{geometric deep learning}~\cite{gdl}, and more recently,
the same approach is being explored in the context of QML~\cite{gqml, cenk_gqml, lie_eqgnn, roy_eqgnn, dong_z2}.

\par To showcase the diversity of applications enabled by hybrid models,
we turn to an important task in high-energy physics (HEP). In particle accelerators,
with the most famous example being the Large Hadron Collider (LHC) at
CERN, vast background signals are generated by scattering experiments.
Among these signals, very weak signatures that could lead to the discovery of new physics
might be contained, perhaps the most notable example so far being the Higgs boson.

\par After several stages in the pipeline, from the collisions and the trigger
systems deciding what signals to be collected, to the reconstruction of tracks,
jets and clusters, in this example we deal with an important analysis task known
as \textit{jet tagging}. Given a set of final-state measurements (four-momenta,
color, charge, flavor, etc) representing jets after parton showering and
hadronization effects, we represent this as a point-cloud and use an appropriate
hybrid Equivariant Quantum Graph Neural Network (EQGNN) to infer whether the
originating particle is a quark or a gluon.

\begin{figure}[!h]
	\center
	\includegraphics[width=0.5\textwidth]{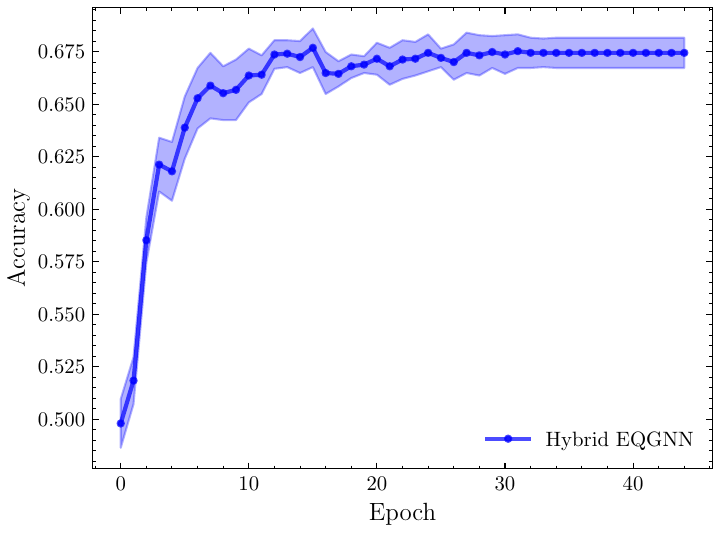}%
	\caption{\label{fig:hybrid_model} Accuracy as a function of training epochs
		obtained executing ten trainings (different initialisation) of the presented
		hybrid model on the quark-gluon dataset. The continuous line is computed as
		median value of the ten accuracies for each epoch and the uncertainties are
		calculated by means of median absolute deviation of the same ten values.}
\end{figure}

Our architectural choice is motivated by the fact that the likelihood of a given
partonic-level process originated from a quark or gluon depends on Lorentz-invariant
matrix elements~\cite{mem_likelihood}. Hence, this is the appropriate symmetry
to encode. However, since the Lorentz group is noncompact,
no finite-dimensional unitary representation exists, making it impossible
to achieve equivariance under the framework proposed by~\cite{gqml}.
As an alternative, motivated by~\cite{lorentznet}, we structure our EQGNN
using universally approximating Lorentz-invariant polynomials~\cite{soledad}.
For simplicity, we include PQCs only in the
\textit{Minkowski dot product attention}~\cite{lorentznet, lie_eqgnn}, $\phi_x$, that acts as:
\begin{equation}
	x^{l}_{i} = x^{l-1}_{i} + c\sum_{j\in \mathcal{N}(i)}\phi_x(m_{ij}^{l})x_j^{l}
\end{equation}
Where $\phi_x$ is a continuous scalar function, modeled by a $4$-layered hybrid
PQC that uses phase encoding followed by trainable $RY$ and $RZ$ rotations
intertwined with entangling layers; $m_{ij}^{l}$ is a Lorentz-invariant message
between particles $i$ and $j$ at layer $l$; and $x_{i}^{l}$ represents the
coordinate embedding (four-momenta in the input layer) for particle $i$ at
layer $l$. In the remaining components of the model, we use standard multi-layered
perceptrons (MLPs). For brevity, here we include only the resulting accuracies,
which go over $10$ trainings with different, randomly initialized weights, and
can be found in figure \ref{fig:hybrid_model}. The detailed architecture together
with a comparison against Lorentznet can be found in~\cite{lie_eqgnn}. We also
refer the interested reader to subsection $4.4$ of~\cite{lorentznet} for a
comparison of equivariant against non-equivariant models on jet tagging. We have
also included a full tutorial in the official \texttt{Qiboedu} repository,
available \href{https://github.com/qiboteam/qiboedu}{here}.

The model was trained over $60$ epochs with a learning rate of $\eta=0.001$
using the \textit{Adam} optimizer, \PyTorch backend and on an ideal
simulator (noiseless and infinite shots). We use the dataset
\textit{Pythia8 Quark and Gluon Jets for Energy Flow}~\cite{quarks_gluons},
which contains two million jets split equally into one million quark jets and
one million gluon jets. These jets resulted from LHC collisions with total center
of mass energy $\sqrt{s} = 14$ TeV and were selected to have transverse momenta
$p_T^{jet}$ between $500$ to $550$ GeV and rapidities $|y^{jet}| < 1.7$.
For our analysis, we randomly picked $N = 12500$ jets and used the first
$10000$ for training, the next $1250$ for validation, and the last $1250$ for
testing. These sets happened to contain $4982$, $658$, and $583$ quark jets,
respectively. We observed that, in practice, this random split happens to be
hard enough to classify for both classical and hybrid models, when their number
of parameters is comparable. It is, thus, a good testbed for models with
different inductive biases.

\subsection{Scaling to larger circuits via tensor network simulation}
\label{sec:tn_scaling}

Classically simulating a quantum circuit exactly is a challenging task as the number of qubits increases.
Statevector simulation scales exponentially with the number of qubits and rapidly becomes impractical beyond a few dozen. 
To address this limitation, tensor network (TN) methods offer an alternative for systems where the entanglement 
structure is constrained.

Many physically-relevant quantum states that present limited entanglement, can be efficiently approximated using tensor 
network methods~\cite{RevModPhys.82.277}. Due to their limited range of correlations, it is possible to efficientlly represent them using low-rank approximations, \emph{e.g.} tensor networks of relatively low bond dimension.
The wavefunction is decomposed into a network of smaller tensors 
interconnected by internal indices of capped dimension, effectively constraining the complexity of the model~\cite{RevModPhys.77.259}.
This approximation enables the efficient classical simulation of quantum systems that would otherwise require exponential
resources. The bond dimension acts as a tunable hyperparameter, balancing representational power and computational efficiency.

\begin{figure}
	\center
	\includegraphics[width=0.5\textwidth]{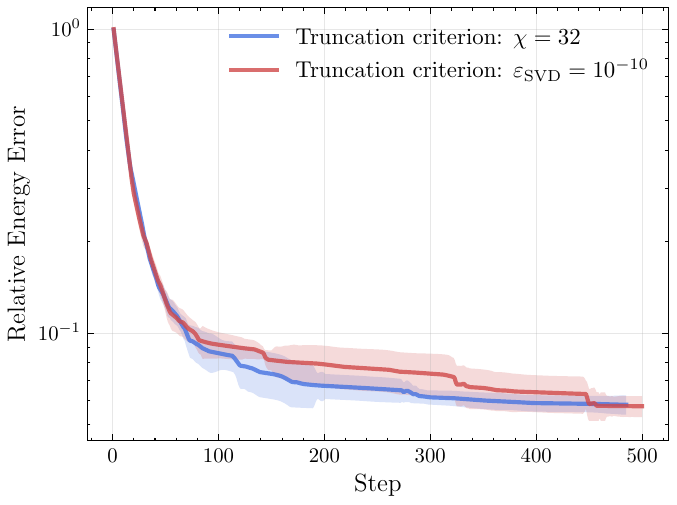}
	\caption{\label{fig:tn_vqe}
  Energy over epochs of the ground state of a $50$ qubits \texttt{XXZ} Hamiltonian prepared by a 
  VQE model trained through a MPS tensor network with standard gradient based optimizers. The 
  training is repeated $10$ times with different initializations, and the solid line and uncertainty
  intervals are obtained from the median and median absolute deviation of the 10 repetitions. 
  The experiment is repeated with two different truncation criteria: a first one with bond dimension
  set to $32$ (blue), and a second one where the truncation is instead controlled by 
  the singolar value decomposition (SVD) cutoff parameter set to $\varepsilon_{\rm SVD}= 10^{-10}$ (red),
  meaning that all singular values smaller than $10^{-10}$ are discarded during MPS truncation.}
\end{figure}

The \Qibo environment provides through \QiboTN some TN backends based on different 
libraries, including \texttt{Quimb}~\cite{quimb}, \texttt{qmatchatea}~\cite{qmatchatea} 
and \texttt{cuTensorNet}. 
Therefore, \QiboTN backends can be used alongside \Qiboml, with the help of the 
\texttt{QuimbJax} differentiation engine introduced in section~\ref{sec:custom_diff}.
This enables scalable quantum simulations and differentiable training workflows
within the familiar \PyTorch or \TensorFlow environments.

\vspace{0.2cm}
As a toy example, we consider the XXZ model with anisotropy in the range $\Delta \in (-1,1]$, described by the Hamiltonian
\begin{equation}
H = \sum_{j=1}^N \,(X_{j}\, X_{j+1} + Y_{j} \, Y_{j+1} + \Delta \, Z_{j} \, Z_{j+1}) \ ,
\end{equation}
where $\{X_{j}, \, Y_{j}, \, Z_{j}\}$ are the usual single-qubit Pauli matrices acting on site $j$, and 
the index equivalency $j = N + 1 \mapsto j = 1$ indicates periodic boundary conditions.
The preparation of eigenstates of the XXZ model on a quantum computer, both variationally 
and exactly, has been the subject of study in several 
recent works~\cite{ho2019efficient,BravoPrieto2020scalingof,sopena_algebraic_2022,ruiz_bethe_2024,FerreiraMartins2025}.
The low-energy spectrum is described by a conformal field theory with central 
charge $c=1$, which means the ground state violates the area law 
logarithmically~\cite{vidal_entanglement_2003,pasquale_calabrese_entanglement_2004}.
This model is exactly solvable via the Bethe ansatz~\cite{Korepin1993kvr,faddeev1996algebraic,gomez_quantum_1996}, 
which allows us to obtain the ground state energy exactly for comparison 
with the results obtained through VQE.
Here, we show as a toy example the training of a $50$-qubit VQE, based on the same 
circuit ansatz of the previous example, to approximate the ground state of the XXZ 
Hamiltonian. This problem, that would be intractable under standard statevector simulations, 
becomes viable with TNs, modulo the issue of vanishing gradients which still curses QML in general.

The obtained results are shown in Figure~\ref{fig:tn_vqe}, where we report the relative error of the estimated energy
as a function of the training epochs, for two different bond dimensions.
The target ground state energy was numerically estimated via a Bethe ansatz for comparison.
We notice that the error decreases over the epochs demonstrating how training such a big model is possible.
However, the energy of the final state seems to still be relatively far from the target ($\sim 6\%$ error).
A more careful choice of the circuit ansatz as well as of optimizer may lead to better results.
Nonetheless, trainability issues are a known problem in QML, and the aim to this example was not to propose a possible solution,
but just to demonstrate the ability to work with very large models.

In the following code snippet, we show how to build and train the VQE model using \Qiboml
with \QiboTN as the simulation backend.

\vspace{0.2cm}
\begin{lstlisting}[language=Python]

from qiboml.models.ansatze import HardwareEfficient
from qiboml.models.decoding import Expectation
from qiboml.operations.differentiation import QuimbJax
from qibo import set_backend
from qibo.hamiltonians import XXZ
import torch
import qiboml.interfaces.pytorch as pt

# Setting the TN backend
# QiboTN is the provider
set_backend(
  "qibotn", 
  platform="quimb", 
  quimb_backend="jax"
)

# Building the quantum model
nqubits = 50
bond_dim = 32

circuit = HardwareEfficient(
  nqubits=nqubits, 
  nlayers=3
)
hamiltonian = XXZ(nqubits, dense=False)

# Using Qiboml API
decoding = Expectation(
    nqubits=nqubits,
    hamiltonian=hamiltonian,
)
# Picking the differentiation engine
diff_engine = QuimbJax(
  ansatz="mps", 
  max_bond_dimension=bond_dim
)

model = QuantumModel(
    circuit_structure=circuit,
    decoding=decoding,
    differentiation=diff_engine
)

# Adopting PyTorch interface
optimizer = torch.optim.Adam(
  model.parameters(), 
  lr=1e-2
)
for epoch in range(100):
    optimizer.zero_grad()
    energy = model()   
    energy.backward()
    optimizer.step()

\end{lstlisting}
\vspace{0.2cm}
Beyond enabling approximation for larger systems, tensor network methods can also extend the tractable circuit 
sizes through a pretraining strategy. As the number of qubits increases, variational quantum algorithms 
often suffer from barren plateaus during optimization. Evidence suggests that 
pretraining parametric quantum circuits using classical TN representations can provide an effective mitigation 
strategy~\cite{Rudolph2023-ud}.

In this approach, the initial quantum circuit is mapped to a tensor network representation with a controlled bond 
dimension, making the training process less susceptible to barren plateaus. Once optimized, the trained TN can be 
converted back into a quantum circuit with improved initial parameters for subsequent fine-tuning, either in 
exact classical simulation or directly on hardware. \Qibo's unified interface facilitates this workflow: a circuit 
can be constructed and pretrained using the \QiboTN backend, then retrieved for further optimization all within 
a single \PyTorch or \TensorFlow environment.

\section{Performance evaluation against other quantum machine learning frameworks}
\label{sec:vs_pennylane}

This section compares the performance of \Qiboml with \PennyLane, a widely
adopted framework for quantum machine learning. We run a series of controlled
experiments across identical settings to assess computational efficiency.

Our benchmark consists in training a VQE model to approximate the ground state of the
following Hamiltonian $H$ 
\begin{equation}
  H = - \sum_{i=1}^{n} Z_i.
\end{equation}
We use a \texttt{HardwareEfficient} ansatz from \Qiboml which is translated into the corresponding circuit in \PennyLane.

\begin{figure}[ht]
  \centering
  \includegraphics[width=0.45\textwidth]{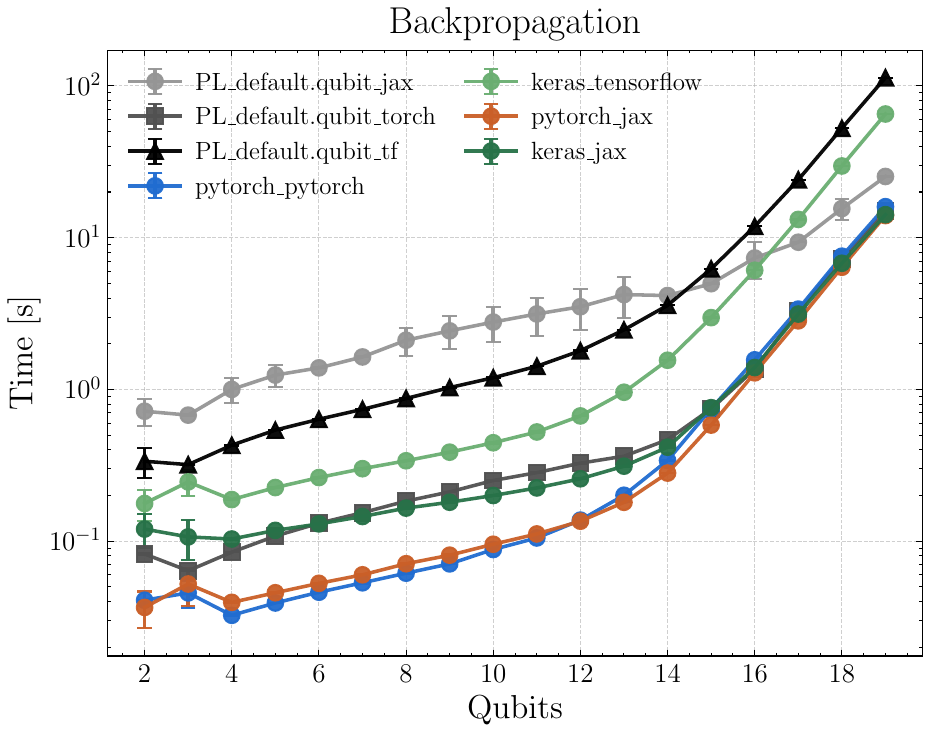}
  \caption{Benchmarking of \Qiboml and \PennyLane using native automatic
  differentiation on a single-thread CPU backend. The plot shows total training
  time for a VQE model as a function of qubit count. In the legend, \PennyLane
  backends are prefixed with \texttt{PL-}\_\texttt{backend}; \Qiboml results are
  labeled as \texttt{interface\_backend}.}
  \label{fig:benchmark_backprop}
\end{figure}

For each system size ($n$), we run 10~\footnote{Prior to the training we compute the gradients once given that some of the
configurations tested are using Just-In-Time compilation, which results in a slower first circuit execution.} 
training epochs and repeat each training 5 times, collecting the total execution time. From these five measurements we report mean and standard
deviation to provide an estimate with uncertainty.

\begin{figure}[ht]
  \centering
  \includegraphics[width=0.45\textwidth]{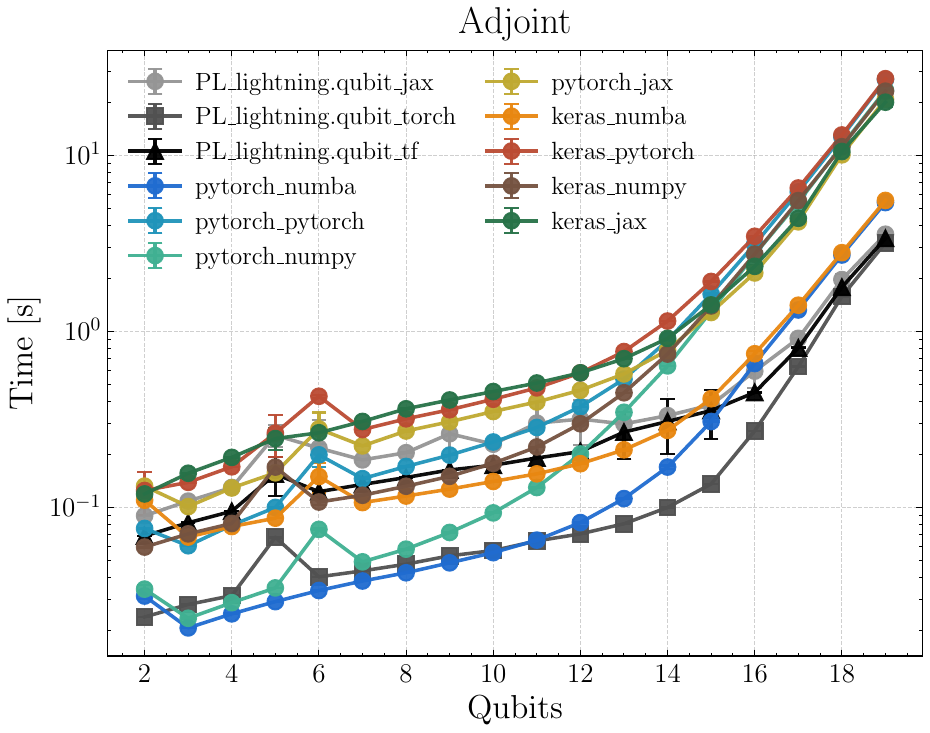}
  \caption{Benchmarking of \Qiboml and \PennyLane with custom adjoint
  differentiation on a single-thread CPU backend. Total VQE training time vs
  qubit count. Legends follow the same convention as in
  Figure~\ref{fig:benchmark_backprop}.}
  \label{fig:benchmark_adjoint}
\end{figure}

We evaluate three differentiation regimes:
\emph{(i)} native automatic differentiation from the host ML framework,
\emph{(ii)} a custom adjoint differentiation engine based on~\citep{jones2020}, and
\emph{(iii)} a custom parameter-shift rule (PSR). In this section we
use both single-threaded and multi-threaded CPUs as well as a GPU environment to address
how the performance of the different configurations change depending on the underline hardware.
For multi-threaded CPU and GPU configurations we focus on \emph{(ii)} since adjoint differentiation
is the main differentation method used when it comes to fast full statevector simulators.
The benchmark is performed on an Intel\textsuperscript{\textregistered} Xeon\textsuperscript{\textregistered} Platinum 8568Y+ Processor
which has 96 threads and on a NVIDIA A40 GPU which has 48 GB of memory. For the multi-threaded benchmark we use 8 threads.

Figure~\ref{fig:benchmark_backprop} reports the single-threaded CPU results with
native automatic differentiation. We then repeat the study using custom adjoint
differentiation. For \PennyLane, we select its optimized C++ adjoint engine as a
representative configuration.

Finally, we compare custom PSR-based differentiation under the same conditions.

\begin{figure}[t]
  \centering
  \includegraphics[width=0.45\textwidth]{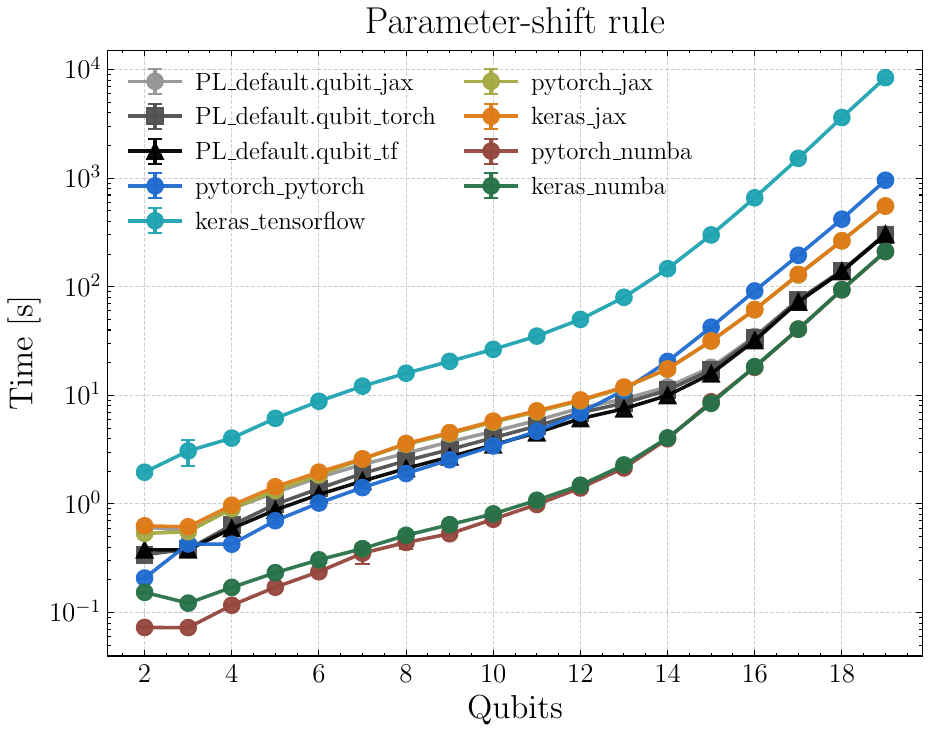}
  \caption{Benchmarking of \Qiboml and \PennyLane using the custom parameter-shift
  rule on a single-thread CPU backend. Total VQE training time vs qubit count.
  Legends follow the same convention as in Figure~\ref{fig:benchmark_backprop}.}
  \label{fig:benchmark_psr}
\end{figure}

\begin{figure*}[t!]
	\centering
	\includegraphics[width=0.49\textwidth]{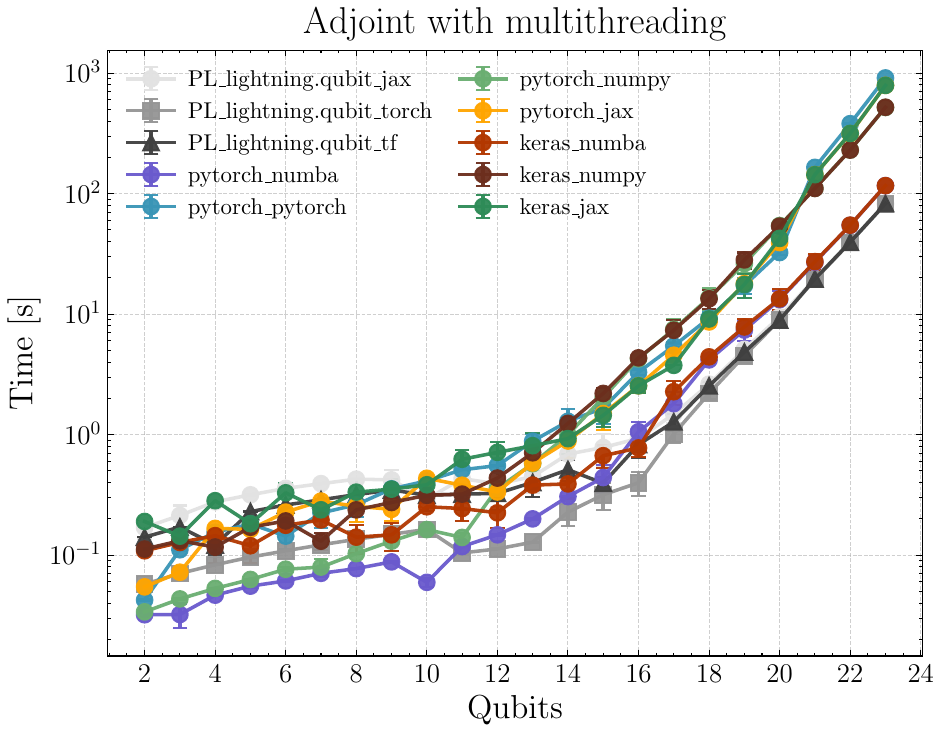}%
	\includegraphics[width=0.49\textwidth]{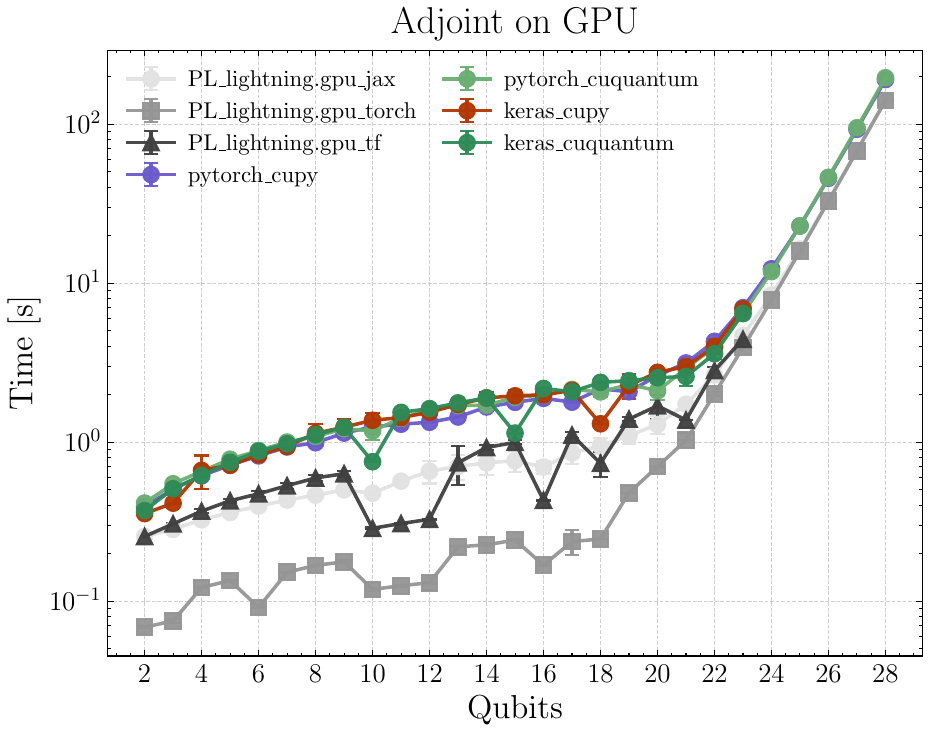}
  \caption{Performance comparison between \Qiboml and \texttt{PennyLane} on CPU running with 8 threads \emph{(left)} and on GPU \emph{(right)}.
  The figure displays the total execution time for VQE model training as a function of qubit count.}
	\label{fig:benchmark_accelerators}
\end{figure*}

The initial evaluation setup involves single-threaded CPU training, as demonstrated in
Figures~\ref{fig:benchmark_adjoint}~-~\ref{fig:benchmark_backprop}.
For adjoint differentiation we observe that although \texttt{Pennylane}'s \texttt{lightning-qubit} is asymptotically
faster using the \texttt{NumbaBackend} from \texttt{Qibojit} \Qiboml performance is reasonably close to \texttt{Pennylane} as shown in Figure~\ref{fig:benchmark_adjoint}.
Using PSR we get a similar behaviour where \Qiboml seems to slightly outperform \texttt{Pennylane} in Figure~\ref{fig:benchmark_psr}.
Instead with backpropagation the two frameworks seems to achieve the same asymptotic performance as shown in Figure~\ref{fig:benchmark_backprop}

In Figure~\ref{fig:benchmark_accelerators} we report the results of the same training running with a CPU with 8 threads and on a GPU using the dedicated backends
of the two frameworks. We are able to run the training for specific configurations up to $28$ qubits. We observe that although
\texttt{PennyLane}'s \texttt{lightning-gpu} is slightly faster, \Qiboml performance is reasonably close to \texttt{PennyLane}. We suppose that the asymptotic overhead between
\Qiboml and \PennyLane is due to the fact that \PennyLane is using directly \texttt{C++} while for \Qiboml we rely on
external libraries such as \texttt{Cupy} to inject \texttt{C++} code in \texttt{Python}. The same applies also for the multithreaded benchmark.

\section{\label{sec:conclusion} Conclusions and outlook}
We presented \Qiboml as a tool for integrating quantum models within
hybrid machine learning workflows. Being part of the \Qibo ecosystem,
\Qiboml inherits all \Qibo features, including the ability to interface
with self-hosted quantum devices through \Qibolab and access to the
characterization and calibration routines provided by \Qibocal. This
heterogeneous environment becomes a playground for researchers and
practitioners, who can easily experiment with different training
setups, ranging from ideal noiseless simulations to real-hardware
training, or explore strategies that benefit from quantum and classical
resources, such as real-time error mitigation and calibration-aware
training.

We provide this tool with the same interfaces as widely used classical
machine learning frameworks, such as \PyTorch and \TensorFlow, to
facilitate the adoption of quantum models in existing classical
pipelines.

Possible directions for future work include involving new hardware
accelerators, such as FPGA boards, to boost performance in dedicated
tasks like real-time operations or recalibration routines. Within the
\Qiboml context, it will also be interesting to explore how classical
and quantum paradigms can support each other: classical for quantum
(for example using modern LLMs or Transformers as support objects), and
quantum for classical, where quantum subroutines may provide utility
within broader classical or hybrid models.

This work represents a further step in two directions: \emph{(i)}
opening quantum computing to a wider audience, and \emph{(ii)} providing
a full-stack resource for researchers and practitioners to explore new
ways to orchestrate quantum and classical resources in the context of
machine learning tasks.

\section{Aknowledgments}
This project is supported by the Quantum Research
Center at the Techonology Innovation Institute (UAE), by the National Research 
Foundation through the National Quantum Computing Hub (Singapore).
This project is also supported by the PNRR MUR project PE0000023-NQSTI (QNIX).
MR acknowledges support from the CERN Doctoral Program through the CERN Quantum 
Technology Initiative during the completion of this work.

%%%%%%%%%%%%%%%%%%%%%%%%%%%%%%%%%%%%%%%%%%%%%%%%%%%%%%%
\bibliography{references}
%%%%%%%%%%%%%%%%%%%%%%%%%%%%%%%%%%%%%%%%%%%%%%%%%%%%%%%

\end{document}